\documentclass[11pt]{amsart}
\usepackage{amsmath, amssymb, amsthm, amscd, graphicx}
\usepackage{epstopdf}
\theoremstyle{plain}
\newtheorem{prop}{Proposition}

\newtheorem{lemm}[prop]{Lemma}

\theoremstyle{definition}

\newtheorem*{defi}{Definition}

\newtheorem*{rema}{Remark}
\newtheorem*{remas}{Remarks}

\newcommand{\CC}{\Gamma}

\newcommand{\Curve}{\Gamma}

\newcommand{\ddd}{\delta}
\newcommand{\DD}{\mathcal D}

\newcommand{\dd}[2]{\delta_3(\Pi_{#2}, \R^{#1})}
\renewcommand{\DD}{\Delta}
\newcommand{\Dist}{\mathrm{Dist}}

\renewcommand{\ggg}{\Gamma}

\newcommand{\pp}{\Pi}

\newcommand{\R}{\mathbb R}

\newcommand{\ZZ}{Z}

\begin{document}

\title{On the 3-distortion of a path}
\author{Pierre Dehornoy}
\address{D\'epartement de Math\'ematiques et Applications,
\'Ecole Normale Superieure, 45 rue d'Ulm, 75005 Paris, France}
\email{dehornoy@clipper.ens.fr}
\thanks{This research was done during a visit at the Institute for
Theoretical Computer Science (ITI) in Prague; the author expresses
his gratitude to Ji\v{r}\'{i} Matou\v{s}ek for his warm hospitality
and his help, as well as to ITI and Charles University for their
support.}
\keywords{distortion, triangle, non-expanding embedding}
\subjclass{68U05}

\maketitle

\begin{abstract}
We prove that, for embeddings of  a path of length $n$ in $\R^2$,
the $3$-distortion is an $\Omega(n^{1/2})$, and that, when embedded
in $\R^d$, the $3$-distortion is an $O(n^{1/(d-1)})$.
\end{abstract}



The general context of this paper is the study of the distortion
that appears when a metric space is embedded into a Euclidean space.
Such a study plays an important role in algorithmic geometry and its
applications. In particular, significant memory gains can be
achieved when a metric space is embedded into a low dimensional
Euclidean space, and, therefore, the study of such embeddings is
directly connected with the construction of efficient computer
representations of (finite) metric spaces. The price to pay for such
memory gains is the inevitable deformations that result from the
embedding, and it is therefore quite important to control them,
typically to understand their asymptotic behaviour when the size of
the metric space increases.

A standard parameter for controlling the deformation is the
distortion, that takes into account pairs of points and compares
their distances in the source and the target spaces---see precise
definition below. The distortion is rather well understood, and, in
particular, precise bounds for its values in the case of general finite metric
spaces are known~\cite{Matousek}.

Now, other parameters may be associated with an embedding naturally.
Typically, for each~$k$, one can introduce the notion of a {\it
$k$-distortion} by taking into account $k$-tuples of points rather
than just pairs, and measuring the way the volume of the associated
polytope is changed. This is what U.\,Feige does in~\cite{Feige} in
order to construct an algorithm minimizing the bandwidth of a graph,
{\it i.e.}, finding a numbering $v_1, ..., v_n$ of the vertices for
which the supremum of $\vert i-j\vert$ over all pairs $(i,j)$ such
that $(v_i,v_j)$ is an edge is as small as possible. The idea
of~\cite{Feige} is to consider volume-respecting embeddings of the
graph into a Euclidean space. The point is to show that, among all
projections of such an embedding on a line, a positive proportion
has a minimal bandwidth of the expected size, and the main step is
to investigate the $k$-distorsion.

Owing to the above applications and connections, understanding
$k$-distortion for every~$k$ seems to be a quite natural goal. Now,
in contrast to the case $k=2$, very little is known so far about
$k$-distortion for $k\ge3$. The aim of this paper is to establish
some results about $3$-distortion, in the most simple case of a
metric space consisting of equidistant points on a line. So, we
denote by $\Pi_n$ the set $\{0, 1, ..., n\}$ equipped with the
distance $d(i,j)=\vert i - j\vert$. Then, for each~$d\ge2$, there
exists a real parameter $\delta_3(\Pi_n,\R^d) \ge1$ that measures
the deformation of triangles when $\Pi_n$ is embedded in~$\R^d$. The
intuition is that, the bigger $\delta_3$, the flatter the
triangles---the precise definition is given in Section~1 below.

As $\R^d$ isometrically embeds in $\R^{d+1}$, the inequality
$\dd{d+1}n \le \dd d n$ immediately follows from the precise
definition, implying in particular $\dd d n \le \dd 2 n$ for $d \ge
3$. The meaning is that, when we have more space, we can more easily
embed with small distortion. For $d = 2$ (the planar case), hence
for every~$d$, it is easy to see that $\dd d n$ is at most linear in
$n$, so the question is to compare $\dd d n$ with the  polynomial
functions $n^\alpha$, $0 < \alpha< 1$. What we do below is to prove
one lower bound result for $d = 2$, and one upper bound result for
$d \ge 2$:

\begin{prop}\label{borneinf}
The $3$-distortion $\dd 2 n$ is an $\Omega(n^{1/2})$.
\end{prop}

\begin{prop}\label{bornesup}
For each fixed $d$, the $3$-distortion $\dd d n$ is an $O(n^{
1/(d-1)})$.
\end{prop}

The results are likely not to be optimal: we conjecture that $\dd2n$
might be an $\Omega(n)$, and that $\dd d n$ might be lower than
polynomial, typically polylogarithmic, for $d \ge 3$. This would
mean that the behaviour of the $3$-distortion radically differs from
the standard distortion which is polynomial in $n$ for each
dimension~$d$.

\section{The $3$-distortion}

Our first task is to make the allusive definitions of the introduction precise.

For $(V, \rho)$ a metric space and $f$ a non-expanding ({\it i.e.},
$1$-Lipschitz) embedding of $V$ into $\R^d$, the distortion~$\DD(f)$
of $f$ is defined to be the supremum of the compression ratio
between the distance of two points in $(V, \rho)$ and that of their
images in $\R^d$:
\begin{equation} \label{EDist}
\DD(f) = \sup\Big\{\frac{\rho(P,Q)}{\Dist(f(P),f(Q))} ; P, Q \in V\Big\}.
\end{equation}
By construction, $\DD(f)$ is at least~$1$, and the larger it is, the
bigger the deformation of distances caused by~$f$.

Let us turn to $k = 3$, {\it i.e.}, let us consider images of
triangles. In the denominator of~\eqref{EDist}, the length of the
segment~$[f(P), f(Q)]$ is replaced with the area of the
triangle~$[f(P), f(Q), f(R)]$. As for the numerator, the area makes
no sense in the source space~$(V, \rho)$, but we observe that, at
least in good cases, $\rho(P,Q)$ is the sup of the lengths
$\Dist(g(P),g(Q))$ for~$g$ a non-expanding embedding of~$V$
to~$\R^d$ (provided $d \ge 1$). This naturally leads to defining
$\rho_3(P, Q, R)$ to be the sup of $\mathrm{Area}([g(P), g(Q),
g(R)])$ for~$g$ a non-expanding embedding of~$V$ to~$\R^d$ (provided
$d \ge 2$), and to defining the $3$-distortion of~$f$ to be
\begin{equation} \label{E3Dist}
\Delta_3(f) = \sup\Big\{\frac{\rho_3(P,Q, R)}{\mathrm{Area}([f(P),f(Q), f(R)])} ;
P, Q, R \in V\Big\}.
\end{equation}
We shall be interested in the minimal possible value of
$\Delta_3(f)$, {\it i.e.}, in the configurations that minimalize the
distortion of triangles. We are thus led to the following notion:

\begin{defi}
The {\it $3$-distortion} $\delta_3(V, \R^d)$ is defined to be the
infimum of $\Delta_3(f)$ over all non-expanding embeddings $f$ of $V$
into $\R^d$.
\end{defi}

The definition for $k$-tuples would be similar, with volume replacing area.

\begin{figure}[htb]
\setlength{\unitlength}{1mm}
\begin{center}
\begin{picture}(110,27)(0, 0)
\put(0,5){\includegraphics[scale=1]{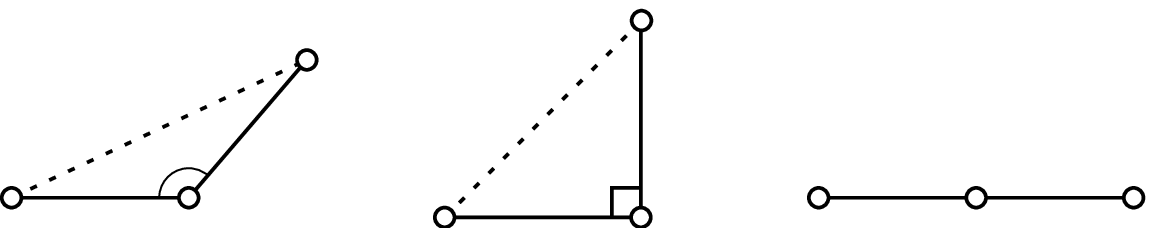}}
\put(14,0){$(i)$}
\put(53,0){$(ii)$}
\put(96,0){$(iii)$}
\put(-6,5){\Small$f(1)$}
\put(20,5){\Small$f(2)$}
\put(31,25){\Small$f(3)$}
\put(9,5){$a$}
\put(26,12){$b$}
\put(18,12){$\theta$}
\put(55,7){$1$}
\put(66,15){$1$}
\end{picture}
\end{center}
\caption{\sf The $3$-distortion of a non-expanding
embedding~$f:\Pi_2\to\R^2$: $(i)$ generic case: the area is
$ab\sin(\theta)/2$, whence $\Delta_3(f)=1/ab\sin\theta$, $(ii)$ an
optimal case: $\Delta_3(f)=1$; $(iii)$ a worst case: the isometrical
embedding; then the image of~$f$ is a flat triangle of area~$0$,
hence $\Delta_3(f)=\infty$.} \label{3dis}
\end{figure}

Figure~\ref{3dis} describes the situation for the graph~$\pp_2$. In
this (very simple) case, there exist embeddings with $3$-distortion
equal to~$1$, namely the ones of Figure~\ref{3dis}$(ii)$, and,
therefore, we find $\delta_3(\pp_2)=1$.

In the general case, we always have $\Delta_3(f) \ge 1$ by
construction, and, the flatter the triangles in the image of~$f$,
the larger $\Delta_3(f)$. For instance, when $f$ is an isometrical
embedding of $\pp_n$ in $\R^d$, all triangles are flat, as in
Figure~\ref{3dis}$(iii)$, and the distortion~$\Delta_3(f)$ is
infinite. Thus the $3$-distortion is a measure of the inevitable
flattening of triangles that occurs when a (large) metric space is
embedded in some fixed Euclidean space: then, it is impossible that
all triples of vertices are embedded so as to form a rectangular
triangle as in Figure~\ref{3dis}$(ii)$, and the question is to
evaluate how far from that one must lie. The reader can check that,
even in the case of embeddings of~$\pp_3$ into~$\R^2$, it is not so
easy to prove that the minimal $3$-distortion
is~$2/\sqrt{3}=1,1547...$, corresponding to a U-shape with
length~$1$ edges and $2\pi/3$~angles, and obtaining an exact value
in the general case of~$\pp_n$ seems out of reach. This contributes
to making asymptotic bounds desirable.

In the specific case of the space~$\Pi_n$, {\it i.e.}, of
$n$~equidistant points at distance~$1$ on the real line, the
definition of $3$-distortion can be given a more simple form. Indeed, if
$g$ is a non-expanding embedding of~$\pp_n$ into $\R^d$, we have
$\mathrm{Dist}(g(i), g(j)) \le  |i-j|$ and therefore, for $i<j<k$,
we find $\mathrm{Area}([g(i), g(j), g(k)]) \le (j-i)(k-j)/2$; on the
other hand, provided $d \ge 2$, we can always find~$g$ such that the
latter inequality is an equality as in Figure~\ref{3dis}$(ii)$.
Hence, for $0 \le i < j < k \le n$, we have
$$\rho_3(i, j, k) = (j-i)(k-j)/2.$$
So, for $f$ a non-expanding embedding of~$\pp_n$ into~$\R^d$,
\eqref{E3Dist} takes the form
\begin{equation} \label{E3Distbis}
\Delta_3(f) =
\sup\Big\{\frac{(j-i)(k-j)/2}{\mathrm{Area}([f(i),f(j), f(k)])} ; 0
\le i < j < k \le n\Big\}.
\end{equation}

In the sequel, we shall forget about embeddings and only work inside
the target space~$\R^d$.

\begin{defi}
A finite sequence of points $(M_0, \dots, M_n)$ in~$\R^d$ is said to
be {\it tame} if, for each~$i$, we have $\Dist(M_i, M_{i+1}) \le 1$.
In this case, we put
\begin{equation} \label{E3DistSuite}
\Delta_3(M_0, \dots, M_n) =
\sup\Big\{\frac{(j-i)(k-j)/2}{\mathrm{Area}([M_i,M_j,M_k])} ;  0 \le
i < j < k \le n\Big\}.
\end{equation}
\end{defi}

If $f$ is an embedding of~$\pp_n$ into~$\R^d$, then the sequence
$(f(0), \dots, f(n))$ is tame, and, conversely, each tame sequence
determines a unique embedding. Now, translating \eqref{E3Distbis}
gives \eqref{E3DistSuite} for $M_i = f(i)$ and the notation is
consistent. Then the $3$-distortion of~$\Pi_n$ can be expressed in
terms of tame sequences of points: for all $n, d$, we have
\begin{equation} \label{E:depart}
\dd d n  = \inf\{\Delta_3(M_0, \dots, M_n) ; (M_0, \dots , M_n)
\text{ a tame sequence in~$\R^d$}\}.
\end{equation}

Thus, from now on, our aim is to study the possible values of the
quantity~$\dd d n$ of~\eqref{E:depart}.

\section{A lower bound in the planar case}

In order to prove Proposition~\ref{borneinf}, we shall consider an
arbitrary tame sequence in~$\R^2$, and prove that some triangle is
much distorted, {\it i.e.}, flattened. To this end we observe that points in convex
position provide a triangle with large $3$-distortion.

Say that a sequence $(P_0, \ldots, P_{m-1})$ of points in the plane
is {\it convex} if the boundary of the convex hull of $\{P_0$,
\dots, $P_{m-1}\}$ is exactly the polygon with vertices $P_0,
\ldots, P_{m-1}$ in this order.

\begin{lemm}\label{Lconvexe}
Assume that $(P_0, \ldots, P_{m-1})$ is a convex sequence with $m\ge
3$. Then there exists $i$ such that the $3$-distortion of the
triangle $P_i P_{i+1} P_{i+2}$---where indices are taken
modulo $m$---is at least $m/(2\pi)$.
\end{lemm}

\begin{proof}
The sum of angles $\angle{P_0P_1P_2} + \angle{P_1P_2P_3} + \ldots +
\angle{P_{m-1}P_0P_1}$ is $(m-2)2\pi$. As all angles are positive
and less than $\pi$, one of them is at least $\frac{m-2}m \pi$. The
$3$-distortion of the corresponding triangle is then at least
$m/(2\pi)$.
\end{proof}

\begin{figure}[htb]
\setlength{\unitlength}{1mm}
\begin{center}
\begin{picture}(40,35)(0, 0)
\put(0,0){\includegraphics[scale=.8]{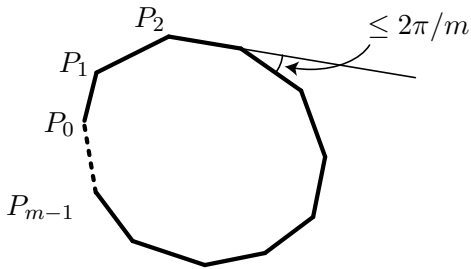}}
\put(-5,18){$P_0$} \put(-3,26){$P_1$} \put(7,32){$P_2$}
\put(38,31){$\le 2\pi/m$} \put(-10,7){$P_{m-1}$}
\end{picture}
\end{center}
\caption{\sf Convex sequence of points} \label{F:convexe}
\end{figure}

\begin{figure}[htb]\label{distdroite}
\setlength{\unitlength}{1mm}
\begin{center}
\begin{picture}(100, 50)(0, 0)
\put(0,0){\includegraphics[scale=.8]{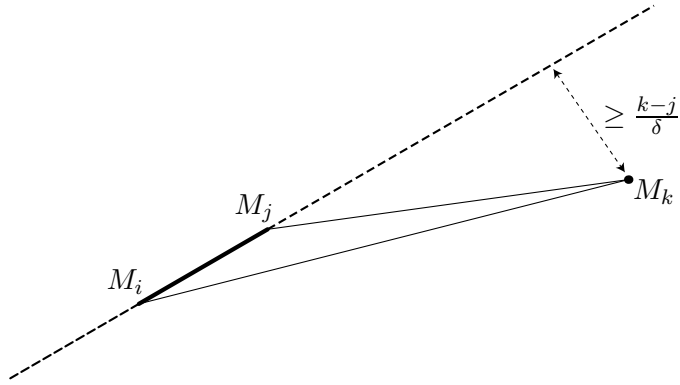}}
\put(13,12){$M_i$} \put(30,22){$M_j$} \put(83,24){$M_k$}
\put(79,34){$\ge\frac{k-j}\ddd$}
\end{picture}
\end{center}
\caption{\sf Minimal distance from the point $M_k$ to the line $(M_iM_j)$}
\label{F:distancemin}
\end{figure}

\begin{lemm}\label{Lsoussuite}
Assume that $(M_0, \dots, M_n)$ is a tame sequence in~$\R^2$, and
that $\ddd$ is an integer greater than or equal to $\Delta_3(M_0,
\dots, M_n)$. Then the sequence $(M_0, M_\ddd, M_{2\ddd}, \ldots,
M_{\lfloor\frac n \ddd\rfloor \ddd})$ is convex.
\end{lemm}

\begin{figure}[htb]
\setlength{\unitlength}{1mm}
\begin{center}
\begin{picture}(80, 34)(0, 0)
\put(0,0){\includegraphics[scale=.8]{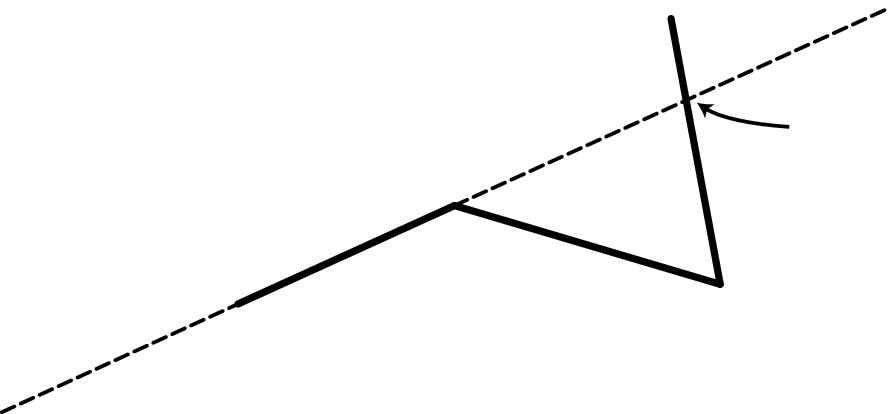}}
\put(15,10){$i\ddd$}
\put(24,18){$(i+1)\ddd$}
\put(59,10){$(i+2)\ddd$}
\put(40,31){$(i+3)\ddd$}
\put(65,22){$problem!$}
\end{picture}
\end{center}
\caption{\sf Four points not in convex position: a problem arises
between $(i+2)\ddd$ and $(i+3)\ddd$}
\label{cas1}
\end{figure}

\begin{figure}[htb]
\setlength{\unitlength}{1mm}
\begin{center}
\begin{picture}(110, 24)(0, 0)
\put(0,0){\includegraphics[scale=.8]{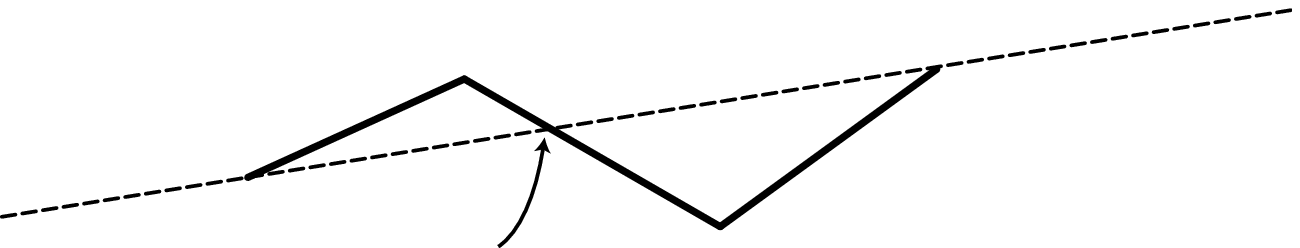}}
\put(16,7){$i\ddd$}
\put(29,16){$(i+1)\ddd$}
\put(59,-2){$(i+2)\ddd$}
\put(77,11){$(i+3)\ddd$}
\put(33,-4){$problem!$}
\end{picture}
\end{center}
\caption{\sf Four points not in ordered convex position: a problem arises
between $(i+1)\ddd$ and $(i+2)\ddd$}
\label{cas2}
\end{figure}

\begin{proof}
Let $\delta_0 := \Delta_3(M_0, \dots, M_n)$. For all $i < j$, we
have $\Dist(M_i, M_j)\le |j-i|$. Since for $k > j$ the area of the
triangle $[M_i, M_j, M_k]$ is at least $\frac{(k-j)(j-i)}{2\ddd_0}$,
hence {\it a fortiori} $\frac{(k-j)(j-i)}{2\ddd}$, the distance
between the point $M_k$ and the line $(M_i M_j)$ is at least
$\frac{k-j}\ddd$ (Figure~\ref{distdroite}). Therefore, for $k\ge
j+\ddd$, the points $M_k$ and $M_{k+1}$ lie on the same side of the
line $(M_i M_j)$: otherwise, the distance between $M_k$ and
$M_{k+1}$ would be at least $2 \, \frac{k-j}\ddd$, contrary to the
tameness hypothesis. Hence, for $k\ge j+\ddd$, the point $M_k$ lies
on the same side of the line $(M_i M_j)$ as $M_{j+\ddd}$.

For a contradiction, assume that, for some $i$,  the sequence
$(M_{i\ddd}$, $M_{(i+1)\ddd}$, $M_{(i+2)\ddd}$, $M_{(i+3)\ddd})$ is
not convex. Then either the four points are not in convex position,
or they are in convex position but they do not appear in the right
order on the border of their convex hull.

In the first case (Figure~\ref{cas1}), one point lies in the convex
hull of the three others. But this contradicts the hypothesis that
adjacent points lie on the same side of each line $(M_{j\ddd}
M_{(j+1)\ddd})$.

In the second case (Figure~\ref{cas2}), the points are in convex position, but the
segment $[M_{(i+1)\ddd}$, $M_{(i+2)\ddd}]$ crosses the line
$(M_{i\ddd}M_{(i+3)\ddd})$. Then there exists $j$ with $(i+1)\ddd
\le j \le (i+2)\ddd$ such that the distance from $M_j$ to
$(M_{i\ddd}M_{(i+3)\ddd})$ is at most $1/2$. The area of the
triangle $[M_{i\ddd}, M_j, M_{(i+3)\ddd}]$ is therefore at most
${3\ddd}/4$. On the other side, by definition of~$\ddd_0$, this area
is at least $((i+3)\ddd-j)(j-i\ddd)/2 \ddd_0$, hence {\it a
fortiori} $((i+3)\ddd-j)(j-i\ddd)/2 \ddd$. Since $(i+1)\ddd\le j\le
(i+2)\ddd$, the latter quantity is at least~$\ddd$, a contradiction.
\end{proof}

\begin{proof}[Proof of Proposition 1]

Let $(M_0, \dots, M_n)$ be a tame sequence in~$\R^2$, and let $\ddd$
be $\lceil\Delta_3(M_0, \dots, M_n)\rceil$. If we have $\lfloor\frac
n \ddd\rfloor < 2$, then we have $\ddd\ge n/2$, hence $\ddd \in
\Omega(n^{1/2})$ {\it a fortiori}. Assume now $\lfloor\frac n
\ddd\rfloor \ge 2$. Then by Lemma~\ref{Lsoussuite}, the sequence
$(M_0, M_{\ddd}, \ldots, M_{\lfloor\frac n \ddd\rfloor \ddd})$ is
convex, and by Lemma~\ref{Lconvexe} there is a triangle whose
distortion is least ${\lfloor\frac n \ddd\rfloor} /{2\pi}$. By
definition, this quantity is at most $\ddd$, hence we have $\ddd \in
\Omega(\frac n \ddd)$. So in any case, $\dd 2 n$ lies
in~$\Omega(n^{1/2})$.
\end{proof}

\begin{rema}
The proof of Lemma~\ref{Lsoussuite} gives many constraints for the
sequence $(M_0,\ldots,$ $M_n)$. Here we use these constraints to
construct a convex subsequence of size $\sqrt{n}$, but it is likely
that larger subsequences with properties slightly weaker than
convexity could be constructed as well. So we think that the result
of Proposition~\ref{borneinf} is not optimal.
\end{rema}

\section{Construction of a $d$-dimensional embedding}

Now we turn to dimension $d$ and we wish to establish the lower
bound result stated as Proposition~\ref{bornesup}. Our aim is to
construct for each $n$ a tame sequence of length $n$ in~$\R^d$ with
a small $3$-distortion, {\it i.e.}, such that all extracted
triangles are not too much flattened.

A natural idea would be to construct the $n$th sequence $(M_{0,n},
\dots, M_{n,n})$ by taking more and more points on a single curve
$\Curve$ of length $1$, and rescaling. But then a small
$3$-distortion would require a complicated curve $\Curve$. Indeed,
assume that $\Curve$ is an immersion of class $C^2$. As $\Curve$ is
compact, the infimum $r_\Curve$ of the radii of the osculating
circles of $\Curve$ is reached at some point, and therefore it is
non-zero. For any $n$, there exists $i$ such that the curvilinear
distance between $M_{i,n}$ and $M_{i+2,n}$ is lower than $2/n$
before rescaling. Then the distances between $M_{i,n}$ and
$M_{i+1,n}$, and between $M_{i+1,n}$ and $M_{i+2,n}$ are lower than
$2/n$ too. Therefore the sine of the angle between the lines
$(M_{i,n}M_{i+1,n})$ and $(M_{i+1,n}M_{i+2,n})$ is at most
$r_\Curve/n$, and the distortion of the triangle
$M_{i,n}M_{i+1,n}M_{i+2,n}$ is at least $n/r_\Curve$. This leads to
a $3$-distortion in $\Omega(n)$ for $(M_{0,n}, \dots, M_{n,n})$. So,
in order to construct sequences of points with small $3$-distortion,
we have either to use curves depending on $n$, or to use a non-$C^2$
curve (typically a fractal curve). In the following construction we
choose the first option.

\begin{proof}[Proof of Proposition 2]
For simplicity, we assume $n=m^{d-1}$ for some $m$. We recursively
construct a family of curves $\ggg_{m,d}$ in $\R^d$, and, on each of
them, we mark $m^{d-1}+1$ points $P_{m, d, 0}, \ldots, P_{m, d,
m^{d-1}}$ in such a way that $\Delta_3(P_{m, d, 0}, \ldots, P_{m, d,
m^{d-1}})$ lies in $O(m)$ for each fixed~$d$.

When $m+1$ points lie at mutual distance 1 on an arc of circle, the
$3$-distortion is in $\Theta(m)$. The idea of our construction is to use
this fact and to recursively put circles one above the others.

Let $\CC_0$ be the sixth of a circle whose radius $r$ will be chosen
later. On $\CC_0$ we put points $P_0, \ldots, P_m$ with regular
angular distance $\frac \pi {3m}$. Then we replace the arc between
$P_i$ and $P_{i+1}$ with a coplanar arc of radius $2r$ lying between
the original arc and the chord connecting $P_i$ to $P_{i+1}$. We
rescale the figure so that the curvilinear coordinate of $P_i$
becomes $i$ for each $i$. We let $\CC_{m,2}$ be the resulting curve
(oriented from $P_0$ to $P_m$) and $P_{m, 2, 0}, \ldots, P_{m, 2,
m}$ be the marked points on $\CC_{m,2}$.

\begin{figure}[htb]
\setlength{\unitlength}{1mm}
\begin{center}
\begin{picture}(130, 100)(0, 0)
\put(0,0){\includegraphics[scale=.8]{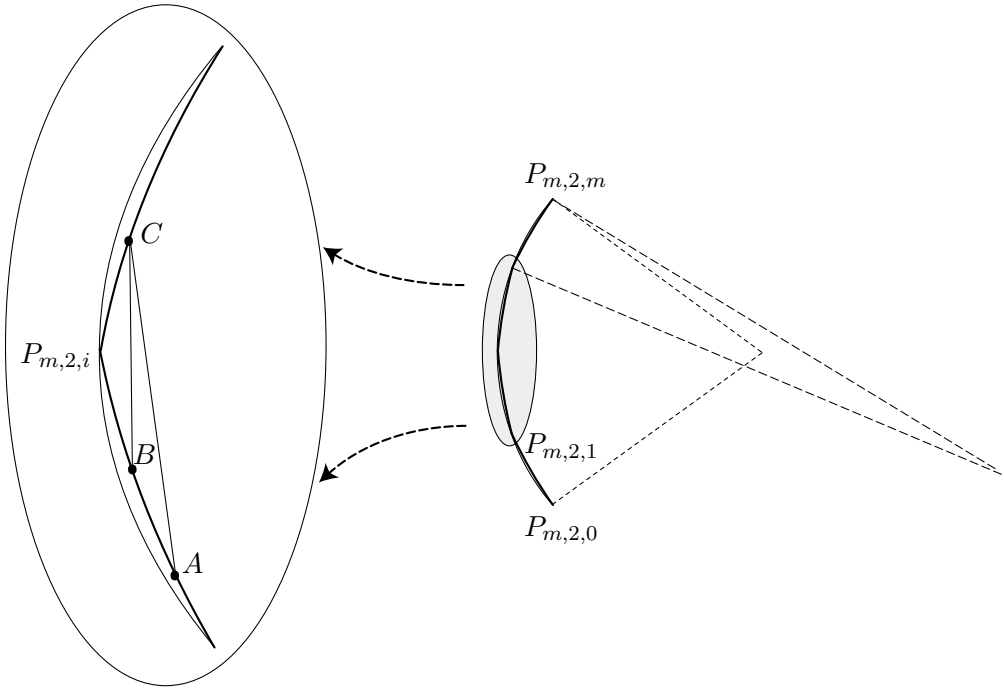}} \put(23.5,15){$A$}
\put(17,29.5){$B$} \put(18,59){$C$} \put(2,43){$P_{m, 2, i}$}
\put(69,20){$P_{m, 2, 0}$} \put(69,31){$P_{m, 2, 1}$}
\put(69,67){$P_{m, 2, m}$}
\end{picture}
\end{center}
\caption{\sf On the right: the curve $\CC_{m,2}$ and the points
$P_{m, 2, 0}, \ldots P_{m, 2, m}$. On the left: three points $A, B,
C$ with at least one $P_{m, 2, i}$ between them yield an angle
$\angle{ABC}\le\pi(1-\frac 1 {6m})$} \label{F:C0}
\end{figure}

The main remark for the proof is that, for all triples $A, B, C$
taken in increasing order on $\CC_{m, 2}$ (not necessarily some
$P_{m, 2, i}$'s) and not all lying on some arc $(P_{m, 2, i}P_{m, 2,
i+1})$, we have $\angle{ABC} \le \pi(1-\frac 1 {6m})$. By construction, the
Euclidean distance between two points of $\CC_{m,2}$ is at least
${3}/ \pi$ times their curvilinear distance, and therefore the
3-distortion of the triangle $ABC$ is in $O(m)$.

The idea for the induction is to add a copy of $\CC_{m, 2}$ between
$P_{m, d-1, i}$ and $P_{m, d-1, i+1}$, orthogonally to the
hyperplane in which $\CC_{m, d-1}$ lies. More precisely, we
construct $\CC_{m,d}$ and $P_{m, d, 0}, \ldots, P_{m, d, m^{d-1}}$
from $\CC_{m,d-1}$ and $P_{m, d-1, 0}, \ldots, P_{m, d-1, m^{d-2}}$
so that the following induction hypothesis is preserved:

$(i)$ $\CC_{m, d}$ is a curve of length $m^{d-1}$ in $\R^{d}$ such
that two points at curvilinear distance $\ell$ lie at euclidian
distance at least $(2/\sqrt{3})^{-d+2}\pi/3\times \ell$;

$(ii)$ If $A, B, C$ are three points that do not all lie
on some arc $(P_{m, d, i}P_{m, d, i+1})$ for any $i$, then the
3-distortion of the triangle $[A, B, C]$ is at most $c_d m$, where
$c_d=(2/\sqrt{3})^{-d+2}\times 6/\pi$.

The induction hypothesis holds for $d=2$.

The construction of $\CC_{m,d}$ is as follows. We identify $\R^d$
with $\R^{d-1}\times \R$, where $\R^{d-1}$ is the space containing
$\CC_{m,d-1}$. Next we work in the cylinder $\ZZ_{m,d-1}$ defined
by~$\CC_{m,d-1}\times \R_+$ with the induced metric. Note that this
cylinder~$\ZZ_{m ,d-1}$ is orthogonal to the hyperplane containing
$\CC_{m,d-1}$. For each $i$ between $0$ and $m^{d-2}-1$, we insert
in~$\ZZ_{m,d-1}$ a rescaled copy of $\CC_{m, 2}$ from $P_{m, d-1,
i}$ to $P_{m, d-1, i+1}$. In this way, we obtain a curve on which
$m^{d-1}+1$ points are marked: the $P_{m, d-1, i}$'s
from~$\CC_{m,d-1}$ plus $m^{d-2}\times (m-1)$ new points between
$P_{m, d-1, i}$ and $P_{m, d-1, i+1}$ for $i=0, \ldots, m^{d-2}-1$.
We denote them by $P_{m, d, 0}, \ldots, P_{m, d, m^{d-1}}$ according
to the linear ordering. We then rescale the figure so that the
curvilinear distance between consecutive points $P_{m, d, i}$'s is
1. We call $\CC_{m, d}$ the resulting curve.

\begin{figure}[htb]
\setlength{\unitlength}{1mm}
\begin{picture}(100, 50)(0, 0)
\put(0,0){\includegraphics[scale=.8]{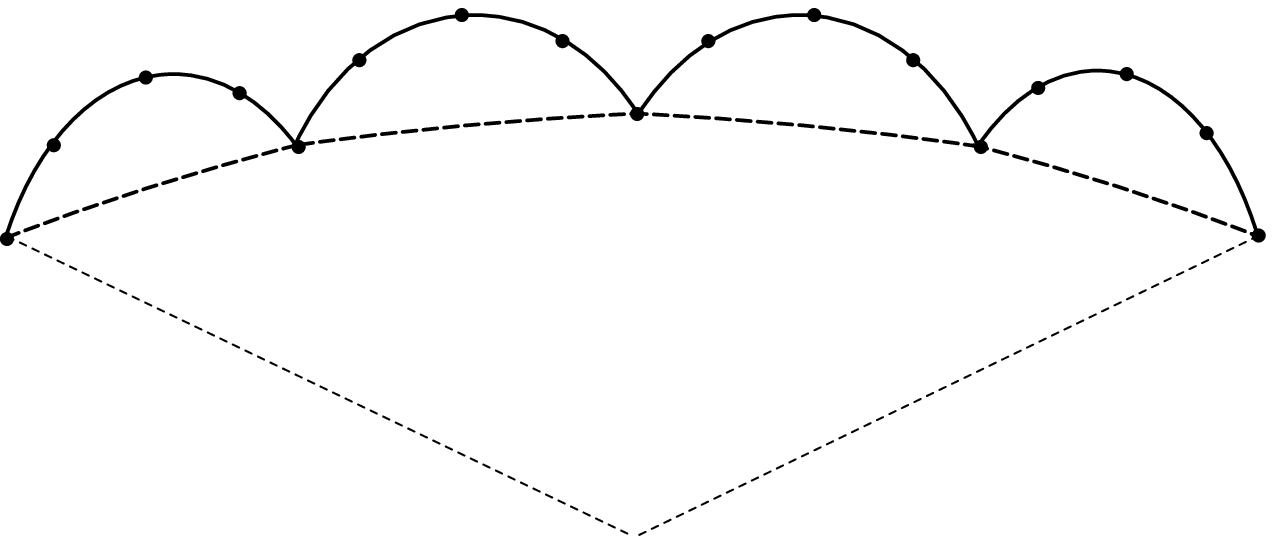}} \put(0,20){$0$}
\put(2,33){$1$} \put(10,39){$2$} \put(24,28.5){$m$}
\put(22,42){$m+1$} \put(49,30){$2m$} \put(100,20){$m^2$}
\end{picture}
\caption{\sf The curve $\CC_{m, 3}$ in the space} \label{F:C3}
\end{figure}

It remains to show that the induction hypothesis is preserved.

For $(i)$, we observe that the angle between any chord of $\ggg_{m,
d}$ and the hyperplane containing $\ggg_{m, d-1}$ is lower than
$\pi/6$. Therefore, when going from $\ggg_{m, d-1}$ to $\ggg_{m,
d}$, no distance is decreased by more than a factor $2/\sqrt{3}$.

For $(ii)$, let $A, B, C$ be three points on $\ggg_{m, d}$ and let $i$
be such that $A$ lies before $P_{m, d, i}$ and $C$ lies after $P_{m, d, i}$
according to the fixed curvilinear ordering.

First case: There exists $j$ such that $A, B, C$ lie between $P_{m,
d, jm}$ and $P_{m, d, (j+1)m}$. This means that $A, B, C$ lie on
some copy of $\CC_{m, 2}$ in~$\ZZ_{m, d-1}$ inserted in the last
step of the inductive construction. In the case of $\CC_{m, 2}$, we
know that the $3$-distortion is at most~$c_2 m$. Here there is an
additional $3$-distortion due to the fact that the copy was made on the
cylinder~$\ZZ_{m, d-1}$. The projection of $\CC_{m,d}$ on $\R^{d-1}$
is $\CC_{m, d-1}$, and not a line as in the $d=2$ case. By induction
hypothesis, the distances on $\CC_{m, d-1}$ (compared with the
Euclidean distances) are not contracted by more than
$(2/\sqrt{3})^{-d+1}\pi/3$, hence the distortion of the triangle
$[A, B, C]$ is bounded by $(2/\sqrt{3})^{-d+1}\pi/3 \times c_2 m \le
c_d m$.

Second case: There exists $j$ such that $A$ lies before~$P_{m, d,
jm}$ and $C$ lies after~$P_{m, d, jm}$. Then, when $A, B, C$ are
projected from $\CC_{m, d}$ on $\CC_{m, d-1}$ along~$\ZZ_{m, d-1}$,
the area of the triangle $[A, B, C]$ decreases by a multiplicative
factor at most $\sqrt 3/2$. By the induction hypothesis the
projection of the triangle has 3-distortion at most $c_{d-1} m$,
therefore the original triangle $[A, B, C]$ has 3-distortion at most
$c_d m$.
\end{proof}

\begin{remas}
$(i)$ The choice of the curve $\CC_{m, 2}$ may look strange, in
particular the choice of an arc of radius $2r$ between $P_{m, i}$
and $P_{m,i+1}$  rather than an arc of radius $r$ or a chord. The
reason is that, in both cases, the key property, namely that the
triangle $[A, B, C]$ has $3$-distortion $O(m)$ if $A, B, C$ do not all
lie on some arc $(P_{d, m, i}P_{d, m, i+1})$, fails. With
arcs of radius $r$, if we take $A, B, C$ close to some $P_{d, m,
i}$, then the 3-distortion of $[A, B, C]$ can be arbitrary large.
With chords, if we take $A, B$ strictly between $P_{d, m, i}$ and
$P_{d, m, i+1}$ and $C$ just after $P_{d, m, i+1}$, then the
3-distortion is not bounded either.

$(ii)$ Our construction uses $d-1$ pairwise orthogonal directions to
draw the curves $\CC_{m, d}$ one above the other. We could use other
fixed directions as well, the point being that the projections
preserve the convexity of the specific patterns we consider.
Alternatively we could replace cylinders by cones, as central
projection also preserves the needed convexity. But it seems
difficult to use more than one cylinder, and therefore more than one
curve, for each new dimension, because no projection preserves the
needed convexity for several sufficiently distinct directions
simultaneously.
\end{remas}

\bibliographystyle{amsplain}

\begin{thebibliography}{4}

\bibitem[1]{Feige} {\sc U. Feige}, Approximating the bandwidth via volume
respecting embeddings, In {\it J. Comput. Sci.}, 60:510-539, (2000).

\bibitem[2]{Matousek} {\sc J. Matou\v{s}ek}, Lectures on Discrete Geometry,
Springer GTM Series, vol. 212, (2001).

\bibitem[3]{Rao} {\sc S. Rao}, Small distortion and volume respecting
embeddings for planar and Euclidean metrics, In {\it Proc. 15th
Annual ACM Symposium on Comput. Geometry}, pages 300--306, (1999).

\bibitem[4]{Vempala} {\sc S. Vempala}, Random projection: a new
approach to VLSI layout, In {\it Proc. 39th IEEE Symposium on
Foundations of Computer Science}, pages 389--395, (1998).

\end{thebibliography}

\end{document}